\documentclass[]{jfm}

\usepackage{amssymb}
\usepackage{amsmath}
\usepackage{epsfig}
\usepackage{epstopdf}
\usepackage{url}
\usepackage{xspace}
\usepackage{color}
\usepackage{txfonts}
\usepackage{graphicx}
\usepackage{bm}
\usepackage{natbib}
\newcommand{\EQ}{\begin{equation}}
\newcommand{\EN}{\end{equation}}
\newcommand{\EQA}{\begin{eqnarray}}
\newcommand{\ENA}{\end{eqnarray}}

\newcommand{\Eq}[1]{Eq.~(\ref{#1})}
\newcommand{\Eqs}[2]{Eqs.~(\ref{#1}) and~(\ref{#2})}
\newcommand{\Eqss}[2]{Eqs.~(\ref{#1}) - (\ref{#2})}

\newcommand{\Fig}[1]{figure~\ref{#1}}
\newcommand{\Tab}[1]{table~\ref{#1}}

\newcommand{\ddt}[1]{\frac{d#1}{dt}}

\def\Rey{\mbox{\rm Re}}
\def\Rep{\mbox{\rm Re}_p}

\newcommand{\Nu}{\mathrm{Nu}}
\def\St{\mbox{\rm St}}
\def\Sh{\mbox{\rm Sh}}
\def\Sc{\mbox{\rm Sc}}
\def\Da{\mbox{\rm Da}}
\def\dam{Damk\"ohler\xspace}

\def\cs{c_{\rm s}}

\def\Xk{\bm{X}^{\rm k}}
\def\Vk{\bm{V}^{\rm k}}

\def\thhh{X}

\newcommand{\Du}{\mathrm{D}}
\newcommand{\del}{\partial}

\newcommand{\uu}{\mbox{\boldmath $u$} {}}

\newcommand{\aaaa}{\mbox{\boldmath $a$} {}}

\newcommand{\ff}{\mbox{\boldmath $f$} {}}

\newcommand{\FF}{\mbox{\boldmath $F$} {}}

\newcommand{\nab}{\mbox{\boldmath $\nabla$} {}}

\newcommand{\SSSS}{\mbox{\boldmath ${S}$} {}}

\newcommand{\yjour}[4]{, #2 {\bf #3}, #4 (#1).}

\newcommand{\yproc}[4]{, (ed. #3), pp. #2. #4 (#1).}
\newcommand{\ybook}[3]{, {\em #2}. #3 (#1).}

\newcommand{\yypr}[4]{, #3, pp. #2. #4 (#1).}

\def\Rey{{\rm Re}}

\def\nab{{\bm{\nabla}}}
\def\ff{{\bm f}}

\newcommand{\bez}{\begin{eqnarray*}}
\newcommand{\eez}{\end{eqnarray*}}
\newcommand{\be}{\begin{equation}}
\newcommand{\beq}{\begin{eqnarray}}
\newcommand{\eeq}{\end{eqnarray}}
\newcommand{\bc}{\begin{center}}
\newcommand{\ec}{\end{center}}

\title{
  The effect of turbulence on mass and heat transfer rates
  of small inertial particles
}
\author{
  Nils Erland L. Haugen$^{1,2}$, 
  Jonas Kr\"uger$^1$, 
  Dhrubaditya Mitra$^3$ and 
  Terese L{\o}v{\aa}s$^1$
}
\affiliation{
  $^1$Department of Energy and Process Engineering, 
  Norwegian University of Science and Technology, 
  Kolbj{\o}rn Hejes vei 1B, NO-7491 Trondheim, Norway\\
  $^2$SINTEF Energy Research, N-7465 Trondheim, Norway\\
  $^3$Nordita, KTH Royal Institute of Technology and Stockholm University, 
  Roslagstullsbacken 23, SE-10691 Stockholm, Sweden
}
\date{}
\begin{document}
\maketitle

\begin{abstract}
The effect of turbulence on the mass and heat transfer between
small heavy inertial particles (HIP) and an embedding fluid is studied.
Two 
effects are identified. The first effect is due to 
the relative velocity between the fluid and the particles, and a 
model for the relative velocity is presented. The second effect
is due to the clustering of particles, where the mass transfer
rate  is inhibited 
due to the rapid depletion of the consumed species inside the
dense particle clusters. 
This last effect is relevant for large \dam
numbers and it may totally control the mass transfer rate
for \dam numbers larger than unity. A model that describes how
this effect should be incorporated into existing particle 
simulation tools is presented.
\end{abstract}

\section{Introduction}
Both in nature and in industrial applications, one regularly finds
small inertial particles embedded in turbulent flows.
By small inertial particles, we mean particles that are smaller than the smallest
scales of the turbulence and have significantly higher material density
than the fluid.
For such particles, there will be momentum exchange between the
particles and the turbulent fluid, and, depending on the conditions, 
there may also be heat
and mass transfer. This is particularly so for
chemically reacting particles, but there are also a large number of
other applications where heat and mass transfer between particles
and fluid are important. Here, the main focus will be on reacting
particles that consume one or more of the
species in the embedding gas through surface reactions.  Relevant
examples are; chemical reactions on the surface of a catalytic
particle, fuel oxidation on the surface of a oxygen carrying particle
in a Chemical Looping Combustion (CLC) reactor, condensation of water
vapor on cloud droplets and combustion or gasification of char.

The presence of turbulence in a fluid will enhance the transport
properties of the flow. This means that the mean-field viscosity,
diffusivity and conductivity may be drastically increased from their
laminar values.  This effect has been studied for many years, and a
large number of different models exist in the literature, such as the
k-$\epsilon$ model (\cite{jones_launder}) and different versions of
the Reynolds Stress Models (e.g. \cite{pope}).  Turbulence may also
modify gas phase combustion, and even though this is somewhat more
complicated, a significant number of models have been developed during
the last decades.  Some examples are the Eddy Dissipation Model
(\cite{EDM}), the Eddy Dissipation Concept (\cite{EDC}) and variations
of Probability Density Function (e.g. \cite{PDF}) models.

With the above knowledge in mind, it is interesting to realize that,
except for the recent work of \cite{kruger}, there is currently no
model describing the effect of turbulence
on the heat and mass transfer of small inertial particles. When a
reacting particle is embedded in a turbulent flow, the turbulence can
potentially influence the mass transfer, and hence the surface
reaction rates in two ways. The first way is through particle clustering,
where particles form dense clusters due to turbulence, and where
the gas phase reactants within the cluster are quickly consumed while
there are no particles that can consume the reactants in the particle
voids outside the clusters. The main effect of the clustering is to
{\it decrease} the overall mass transfer rate. The second way
turbulence influence the mass transfer rate is by increasing the 
mean velocity difference
between the particle and the gas. This effect will {\it increase} the
mass transfer rate.

The same two effects are also active for the heat transfer. The 
similarity between heat and mass transfer can be seen by
considering the expressions for the transfer coefficients of mass 
\EQ
\label{kappa}
\kappa=\frac{\Sh D}{d_p}
\EN
and heat
\EQ
\kappa_{\rm th}=\frac{\Nu D_{\rm th}}{d_p},
\EN
where 
$d_p$ is the particle radius,
$\Sh$ and $\Nu$ are the Sherwood and Nusselt numbers and
$D$ and $D_{\rm th}$ are the mass and thermal diffusivities.
For single spherical particles in flows with low and medium 
particle Reynolds numbers, the Sherwood and Nusselt numbers can be
approximated by the empirical expressions of \cite{ranz_marshall}
\EQA
\label{sherwood}
\Sh_{\rm RM}&=&2+0.69\Rey_p^{1/2}\Sc^{1/3}\\
\Nu_{\rm RM}&=&2+0.69\Rey_p^{1/2}\Pr^{1/3}\nonumber.
\ENA

A well known example where reacting particles are consumed in a
turbulent fluid is the case of pulverized 
coal combustion, where turbulence influences the process in
several ways that are understood to varying degrees.
The combustion of coal can be divided into four separate processes; 
1) drying,
2) devolatilization, 
3) combustion of volatiles and 
4) burnout of the remaining char. 
Processes 1 and 2 involve the evaporation of fluids and thermal cracking of
hydrocarbons, while process 3 involves homogeneous reactions. 
In process 4, gas phase species diffuse to the particle surface and 
react with the solid carbon. 
This happens via adsorption of e.g. an oxygen radical to a carbon
site on the particle surface
and a subsequent desorption of carbon monoxide into the gas phase. This
makes process 4 dominated by heterogeneous chemical reactions.
Many published studies utilize RANS based simulation tools 
that describe simulations of pulverized coal conversion 
in the form of combustion or gasification with 
an Eulerian-Eulerian approach 
(\cite{gao_2004} and \cite{zhang_2005})
or a Lagrangian-Eulerian approach
(\cite{silaen_wang_10,vascellari_etal_14,vascellari_etal_15,klimanek_etal_15,
chen_etal_12,chen_etal_00}). However, none of these papers take the effect of turbulence
on the heterogeneous char reactions into account. To the knowledge of
the authors, the only studies where account is made for this effect are the papers
of \cite{luo_2012, brosh_2014, brosh_2015} and \cite{hara_2015} 
where the Direct Numerical Simulations (DNS) approach is used.
In a DNS,
all turbulence scales are explicitly resolved on the computational grid, such
that the effect of turbulence is implicitly accounted for. 
However, the DNS approach is extremely costly and can therefore only be
used for small simulation domains. For simulations of large scale
applications, the RANS or LES based simulation tools will therefore 
be the only applicable tools for the foreseeable future.

In the current paper, the same framework as was developed by 
\cite{kruger} has been used and extended.
The aim of the paper is to identify the effect of turbulence on 
the mass and heat transfer of solid particles, and to develop models
that describe this effect for all Damk{\"o}hler numbers.


\section{Mathematical model and implementation}
In the current work, the so called point-particle direct numerical
simulation (PP-DNS) approach is used. Here, the turbulent fluid itself
is solved with the direct numerical simulation (DNS) methodology,
where all turbulent scales are resolved and no modelling is
needed. The particles are however not resolved, but rather treated
as point particles where the fluid-particle momentum, mass and heat
interactions are modelled.  The point particle approach is a
simplification that relies heavily on the quality of the models. The
alternative approach, which is to resolve the particles and their
boundary layer, is extremely CPU intensive and can currently not be
done for more than a few hundred particles, even on the largest
computers (\cite{deen_kuipers_2014}).

A number of simplifications are made in this paper. This has been done
in order to make the simulations less CPU intensive, and, even more
importantly, to isolate the dominating physical mechanisms. The
particles are considered to be ever lasting, i.e. they are not
consumed.  The reaction on the particle surface is converting reactant
A to product B;
\EQ
A\rightarrow B
\EN
isothermally, i.e.; there is no production or consumption of
heat, such that only the mass transfer effect is considered.
As explained above, the effect on the heat transfer rate will 
be similar to the effect on the mass transfer rate.
As reactant A is converted product B,
the thermodynamical and
transport properties are not changed.

\subsection{Fluid equations}
The equations determining the motion of
the carrier fluid is give by the continuity equation
\EQ
\label{cont}
\frac{\del \rho}{\del t}+ \nabla\cdot( \rho \uu)=0, 
\EN
and the Navier--Stokes equation
\EQ
\rho \frac{\Du \uu}{\Du t}=-\nab P+\nab \cdot (2\mu \SSSS)+\rho \ff + \FF.
\label{eq:momentum}
\EN
Here, $\rho$, $\uu$, $\mu=\rho\nu$ and $\nu$ are the density, velocity and 
dynamic and kinematic viscosities of the carrier fluid, respectively. 
The pressure $P$ and the density $\rho$ are related by the 
isothermal sound speed $\cs$, i.e., 
\EQ
P=\cs^2\rho,
\EN
while the trace-less rate of strain tensor is given by
\EQ
\SSSS = \frac{1}{2}\left( \nab \uu + (\nab \uu)^T\right)
-\frac{1}{3}\nab \cdot \uu.
\EN
Kinetic energy is injected into the simulation box through the forcing 
function $f$, which is solenoidal and non-helical and injects energy and 
momentum perpendicular to a random wave vector whose direction changes 
every time-step  
\citep{haugen_etal12,kruger}. Similar kinds of 
forcing has also previously been used for particle laden
flows by other groups \citep{bec07}.
The energy injection rate is maintained at a level such that
the maximum Mach number is always below 0.5.
The domain is cubic with periodic boundaries in all directions.
The momentum exchange term, $\FF$, is chosen to conserve momentum between the 
fluid and the solid particles, i.e.,
\begin{equation}
\FF =- \frac{1}{V_{\rm cell}}\sum_{\rm k}m^k\aaaa^k
\label{eq:FF}
\end{equation}
when $V_{\rm cell}$ is the volume of the grid cell of interest and
$m^k$ and $\aaaa^k$ are the mass and acceleration (due to fluid drag)
of the k'th particle within the grid cell.

The equation of motion of the reactant has the well-known
advection-reaction-diffusion form:
\EQ
\label{scalar}
\frac{\del \thhh}{\del t}+\nabla\cdot(\thhh \uu) 
=  D\bar{M}_{\rm c}\nabla\cdot (\nabla \thhh)+\tilde{R} \/,
\EN
where $\thhh$, $\bar{M}_{\rm c}$ and $D$   are the mole fraction, 
the mean molar mass and the diffusivity of the reactant, respectively. 
The last term in \Eq{scalar}, $\tilde{R}$, is the sink term due to 
the gas-solid reactions on the surface of the solid particles.

\subsection{Particle equations}
The $N_p$ particles that are embedded in the flow are treated as point
particles, which means that they are assumed to be significantly smaller
than the viscous scale of the fluid and the diffusive scale of the reactant.
The motion of the k'th particle is described by the equations for position 
\EQ
\ddt{\Xk} = \Vk
\EN
and velocity 
\EQ
\ddt{\Vk} = \aaaa^k
\EN
when the particle acceleration due to fluid drag is given by
$\aaaa^k=\frac{1}{\tau} \left[\uu(\Xk) - \Vk \right]$. Note that gravity 
is neglected in this work.
The particle response time is given by (\cite{schiller1933}) 
\EQ
\tau=\frac{\tau_{\rm St}}{1+f_c}
\EN
when $\tau_{\rm St}=S d_p^2/18\nu$ is the Stokes time,
$f_c=0.15\Rep^{0.687}$ is a Reynolds number correction term to the classical 
Stokes time, 
$S=\rho_p/\rho$ is the density ratio, $\rho_p$ is
the material density of the particles,
\EQ 
\label{Rep}
\Rep=\frac{|\uu(\Xk)-\Vk|d_p}{\nu} =\frac{u_{\rm rel}d_p}{\nu }
\EN
is the particle Reynolds number and $d_p$ is the particle diameter.

\subsection{Surface reactions}
Let us now model the reactive term. 
We assume that the reactions are limited to the surface of the particles
and that the reactions are diffusion controlled, i.e. that all reactant that
reaches the particle surface is consumed immediately\footnote{It is possible
to relax the assumption of diffusion controlled reactions by also accounting
for chemical kinetics at the particle surface, see \cite{kruger}.}. 
The reactive term can then be written as 
\EQ
\label{moltheta}
\tilde{R}=\frac{1}{V_{\rm cell}}\sum_{\rm k}A_p^k \kappa X^K_\infty
\EN
where $A_p=4\pi r_p^2$ is the external surface area
of the particle, the mass transfer coefficient is given by 
\begin{equation}
\kappa=\frac{D\Sh}{d_p}
\label{eq:kappa}
\end{equation}
and $\Sh$ is the Sherwood number.

To couple the reactive particle with the continuum equations
we use the following prescription; for the ${\rm k}$-th particle, 
which is at position $\Xk$,  we set 
\begin{equation}
\thhh_{\infty}^k =\thhh(\Xk),
\label{eq:th}
\end{equation}
i.e.; the far field reactant mole fraction is set equal to the 
reactant mole fraction of the fluid cell where the particle is. 
In the current work, the particle Sherwood number is determined by the
expression of \cite{ranz_marshall} (see \Eq{sherwood} in the introduction),
which is in contrast to the work of \cite{kruger} where
the Sherwood number was set to a constant value of 2, which 
corresponds to the Sherwood number in a quiescent flow.
The particle Reynolds number is given by \Eq{Rep} and
the Schmidt number, $\Sc=\nu/D$, is the ratio of the fluid 
viscosity and the mass diffusivity.

\subsection{The reactant consumption rate}
It is useful to define a reactant consumption rate as
\EQ
\alpha=-\overline{
  \left(
  \frac{\tilde{R}}{X_\infty}
  \right)
}
=\overline{n_p A_p \kappa},
\EN
when $\overline{O}$ represents the volume average of flow property $O$ 
and $n_p$ is the particle number density.
If everything is assumed to be homogeneously distributed over the volume,
the reactant consumption rate is given by
\EQ
\label{alpha_hom}
\alpha_{\rm hom}=n_p A_p \kappa=n_pA_p\frac{\Sh D}{d_p}
\EN
for a given particle size and number density. 

In many RANS based simulation tools, where the local 
fluid velocity is not resolved, it is common to neglect the relative 
velocity difference between the turbulent eddies and the particles.
This implies that $\Sh= 2$.
Since the effect of particle clustering is also neglected in such 
models,
the modelled reactant consumption rate becomes;
\EQ
\label{alpha_norm2}
\alpha_{{\rm Sh, Da}}
=\lim_{{\rm Sh} \rightarrow 2, {\rm Da} \rightarrow 0}\alpha
=n_pA_p\frac{2D}{d_p}.
\EN
In the following, $\alpha_{{\rm Sh, Da}}$
will be used for normalization. 

It is useful to define the
\dam number, which is the ratio of the typical turbulent and chemical
time scales, as
\EQ
\label{dam}
\Da=\frac{\tau_L}{\tau_c}
\EN
where $\tau_L=L/u_{\rm rms}$ is the integral time scale of the turbulence,
$L$ is the turbulent forcing scale, $u_{\rm rms}$ is the root-mean-square
turbulent velocity and
the chemical time scale is
\EQ
\label{tauc}
\tau_c=1/\alpha_{\rm Sh, Da}.
\EN

Particles in a turbulent flow field will tend to form clusters with
higher particle number density than the average
\citep{Squires1991,Eaton1994,Toschi2009,Wood2005}. 
If the chemical time scale is short compared 
to the life-time of the clusters, the reactant concentration within the 
clusters will be much lower than outside the clusters. On the other hand,
if the particle number density is low, the particle clusters will not have enough
time to consume a significant fraction of the reactant during the life-time
of the cluster, and hence, the reactant concentration will be roughly the
same inside as it is outside the clusters. By assuming that the life-time of the
clusters is of the order of the turbulent time scale, it is clear that 
the reactant concentration of particle flows with low \dam number 
will behave as if the particles 
were homogeneously distributed over the volume, i.e.; for small \dam
numbers there is no effect of particle clustering on the reactant consumption. 

From \Eqss{alpha_hom}{tauc} it can be deduced that for the 
homogeneous case, and then also for all cases with low \dam numbers,
the reactant consumption rate will scale linearly with the \dam number
for a given turbulent flow field, such that
\EQ
\alpha_{\rm hom}=\frac{\Da}{\tau_L}\frac{\Sh}{2}.
\EN

When relaxing the restriction to small \dam numbers, the effect of 
particle clustering eventually comes into play.
\cite{kruger} have shown that
the reactant consumption rate is given by
\EQ
\alpha=\frac{\alpha_c\alpha_{\rm hom}}{\alpha_c+\alpha_{\rm hom}}
\EN
when $\alpha_c$ is a cluster
dependent decay rate. (Note that since Kruger et al. assumed the Sherwood
number to be 2, their $\alpha_{\rm hom}$ equals our 
$\alpha_{{\rm Sh},{\rm Da}}=\Da / \tau_L$.)
From this expression,  
the following normalized reactant consumption
rate is found 
\EQ
\label{alpha_tilde}
\tilde{\alpha}_{\rm Sh}
=\frac{\alpha_{\rm Sh}}{\alpha_{\rm Sh, Da}}
=\frac{\alpha_c\tau_L}{\alpha_c\tau_L+\Da\Sh/2}\frac{\Sh}{2}.
\EN
when $\Sh$ is given by \Eq{sherwood} and 
the corresponding relative velocity between the particle and the fluid
is determined by a model (which will be obtained in
the next subsection).
For diffusion controlled reactions, the modified reaction decay
rate, as given by \Eq{alpha_tilde}, 
is a measure of the relative modification to the mass transfer rate
due to the effect of turbulence. This means that a 
modified Sherwood number can now be defined that accounts for the 
effect of turbulence;
\EQ
\label{Sh_mod}
\Sh_{\rm mod}=2\tilde{\alpha}.
\EN
In the limit of small \dam numbers, this expression reduces to
$\Sh_{\rm mod}=\Sh$, as expected.

By employing the modified Sherwood number given by \Eq{Sh_mod}, one can now use 
the common expression for the reactant consumption rate, as
given by \Eq{alpha_hom}, to find the real reactant consumption rate. 
In most cases, however, one needs the particle conversion
rate $\dot{n}_{\rm reac}$ for individual particles, which is closely
connected to the reactant decay rate. 
For diffusion controlled mass transfer, the particle conversion
rate is given by $\dot{n}_{\rm reac}=-\kappa X_\infty C_g$, where $C_g$
is the molar concentration of the gas phase and the mass
transfer coefficient is now found by using the modified 
Sherwood number (as given by \Eq{Sh_mod}) in;
\Eq{kappa}
\EQ
\label{kappa2}
\kappa=\frac{D\Sh_{\rm mod}}{d_p}.
\EN
In many applications, the mass transfer rate is not purely 
diffusion controlled. This can be accounted for by including the 
effect of reaction kinetics at the particle surface.
The corresponding particle 
conversion rate can then be expressed as \citep{kruger}
\EQ
\dot{n}_{\rm reac}=-\frac{\lambda \kappa}{\lambda + \kappa} X_\infty C_g,
\EN
where $\lambda$ is the surface specific molar conversion rate.
Since the reaction kinetics is only dependent on the conditions
at the particle surface, the surface specific molar conversion 
rate is not affected by the turbulence. This is, as we have already
seen, not the case for the mass transfer coefficient, which is
now given by \Eq{kappa2}. 
In this way, all the common machinery for calculating
particle reaction rates can still be used since the effects of the
turbulence are incorporated into the modified Sherwood number.

\section{Results}
In all of the following, statistically stationary homogeneous and isotropic 
turbulence is considered. The Reynolds number 
is varied by changing the domain size while maintaining constant viscosity and
turbulent intensity. 
The \dam number is varied by changing the number density of
particles, while keeping everything else the same.
All relevant simulations are listed in \Tab{runs}.
\begin{table}
\caption{Summary of the simulations. The fluid density is unity while the
Schmidt number is 0.2 and the viscosity is $2\times 10^{-4}$~m$^2$/s for all
the simulations. 
For every simulation listed here, a range of identical simulations
with different \dam numbers have been performed.}
\centering
\setlength{\tabcolsep}{3pt}
\begin{tabular}{l c c c c c c c c c c c c c}
\hline
Label  & $L$ (m) & $N_{\rm grid}$ & $d_p$ & $\rho_p$& Re & Sh & St &$\tau_L$& $\alpha_c$&$\alpha_c \tau_L St/Sh$\\
\hline
1A     & $\pi/2$ &  $64^3$      &$3.4\times 10^{-3}$ & 50     &   80&2.5 &1.0 &1.6&0.9   &0.63\\
2A     & $2\pi$  &  $128^3$     &$19\times 10^{-3}$  & 50     &  400&2.8 &1.0 & 5 &0.23  &0.43\\
3A     & $8\pi$  &  $256^3$     &$11\times 10^{-3}$  & 500    & 2200&2.8 &1.0 &15 &0.07  &0.41\\
\hline
2AB    & $2\pi$  &  $128^3$     &$19\times 10^{-3}$  & 25     &  400&2.7 &0.5 & 5 &0.26  &0.25\\
3AB    & $8\pi$  &  $256^3$     &$11\times 10^{-3}$  & 250    & 2200&2.6 &0.5 &15 &0.09  &0.26\\
\hline                                                                                       
2B     & $2\pi$  &  $128^3$     &$11\times 10^{-3}$  & 50     &  400&2.5 &0.3 & 5 &0.21  &0.13\\
3B     & $8\pi$  &  $256^3$     &$11\times 10^{-3}$  & 150    & 2200&2.6 &0.3 &15 &0.09  &0.18\\
\hline                                                                                       
2C     & $2\pi$  &  $128^3$     &$19\times 10^{-3}$  & 5      &  400&2.4 &0.1 & 5 &0.55  &0.12\\
3C     & $8\pi$  &  $256^3$     &$11\times 10^{-3}$  & 50     & 2200&2.4 &0.1 &15 &0.20  &0.13\\
\hline
2D     & $2\pi$  &  $128^3$     &$19\times 10^{-3}$  & 1.5    &  400&2.3 &0.03 & 5 &1.20 &0.08\\
3D     & $8\pi$  &  $256^3$     &$11\times 10^{-3}$  & 16     & 2200&2.3 &0.03 &15 &0.45 &0.10\\
\hline
2E     & $2\pi$  &  $128^3$     &$19\times 10^{-3}$  & 0.5    &  400&2.2 &0.001& 5 &4.10 &0.10\\
\end{tabular}
\label{runs}
\end{table}

\subsection{The mean relative particle velocity}
In order to predict a representative value of the particle Sherwood number 
from \Eq{sherwood}, the particle Reynolds number $\Rey_p$ is required. 
From \Eq{Rep} it is clear that this also requires 
the relative particle velocity $u_{\rm rel}$, which will be found in this 
subsection. 

Given a particle with a response time that equals the Stokes time;
\EQ
\tau_p=\frac{Sd_p^2}{18\nu},
\EN
such that $\tau_k<\tau_p<\tau_L$, where $\tau_k$ is the Kolmogorov time
scale and $\tau_L$ is the integral time scale. 
With respect to the particle-turbulence interactions, the turbulent
power spectrum may be divided into three distinct regimes, based
on the relation between the particle response time and the 
turbulent eddy turnover time $\tau_{\rm eddy}$.
The first regime is defined as the section of the turbulent power
spectrum where the turbulent 
eddies have turnover times that are much larger than
the response time of the particles, i.e. where $\tau_{\rm eddy} \gg \tau_p$.
All the turbulent eddies in this regime will see the particles as passive
tracers, which follow the fluid perfectly. 
I.e., there will be no relative velocity between the particles and the 
eddies. 
The third
regime is defined as the part of the power spectrum where the turbulent
eddies have much shorter time scales than the particles, i.e. where 
$\tau_{\rm eddy} \ll \tau_p$. The eddies in regime three will see the particles as
heavy bullets that move in straight lines, without being affected by
the motion of the eddies. Hence, the velocity of these eddies will contribute
to the relative particle-fluid velocity. 
The second regime is now defined as the
relatively thin band in-between regimes one and three, where 
$\tau_{\rm eddy} \approx \tau_p$. 
These
are the eddies that are responsible for particle clustering, since they
are able to accelerate the particles to a level where they are thrown out
of the eddy due to their inertia. 
In the following, we will refer to a typical eddy in regime two
as a {\it resonant eddy}, and we define the scale of this eddy as $\ell$.
The resonant eddies are identified by their time scale, $\tau_\ell$, which 
is of the order of the particle response time, $\tau_p$. 
For convenience,
we set the two time scales equal, such that 
\EQ
\label{res_time}
\tau_\ell=\tau_p.
\EN

Based on the definitions above, it is clear that the largest 
turbulent eddies that 
yield a relative velocity between the fluid and the particles, 
are the resonant eddies. 
By assuming Kolmogorov scaling, the velocity of 
the resonant eddies is known to be 
$u_\ell=u_{\rm rms}(\ell/L)^{1/3}$, 
which can be combined
with the above expression for the time scales to yield 
\EQ
\label{kl}
k_\ell=k_L\St^{-3/2}
\EN
when the particle Stokes number is defined as
\EQ
\St=\frac{\tau_p}{\tau_L}
\EN
and $k_\ell=2\pi/l$ and $k_L=2\pi/L$ are the wave-numbers of the resonant
eddies and 
the integral scale, respectively. In obtaining \Eq{kl}, it has also
been used that the turnover time of the resonant eddies is $\tau_\ell=l/u_\ell$,
while that of the integral scale eddies
is $\tau_L=L/u_{\rm rms}$.
\begin{figure}
\centering\includegraphics[width=.45\textwidth]{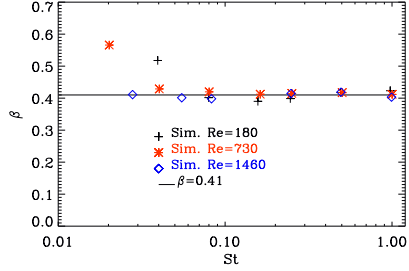}
\caption{
The parameter $\beta$, 
relating the relative particle
velocity to the subscale velocity as defined in \Eq{urel}, is shown
as a function of Stokes number.}
\label{fig:beta}
\end{figure}
\begin{figure}
\centering\includegraphics[width=.45\textwidth]{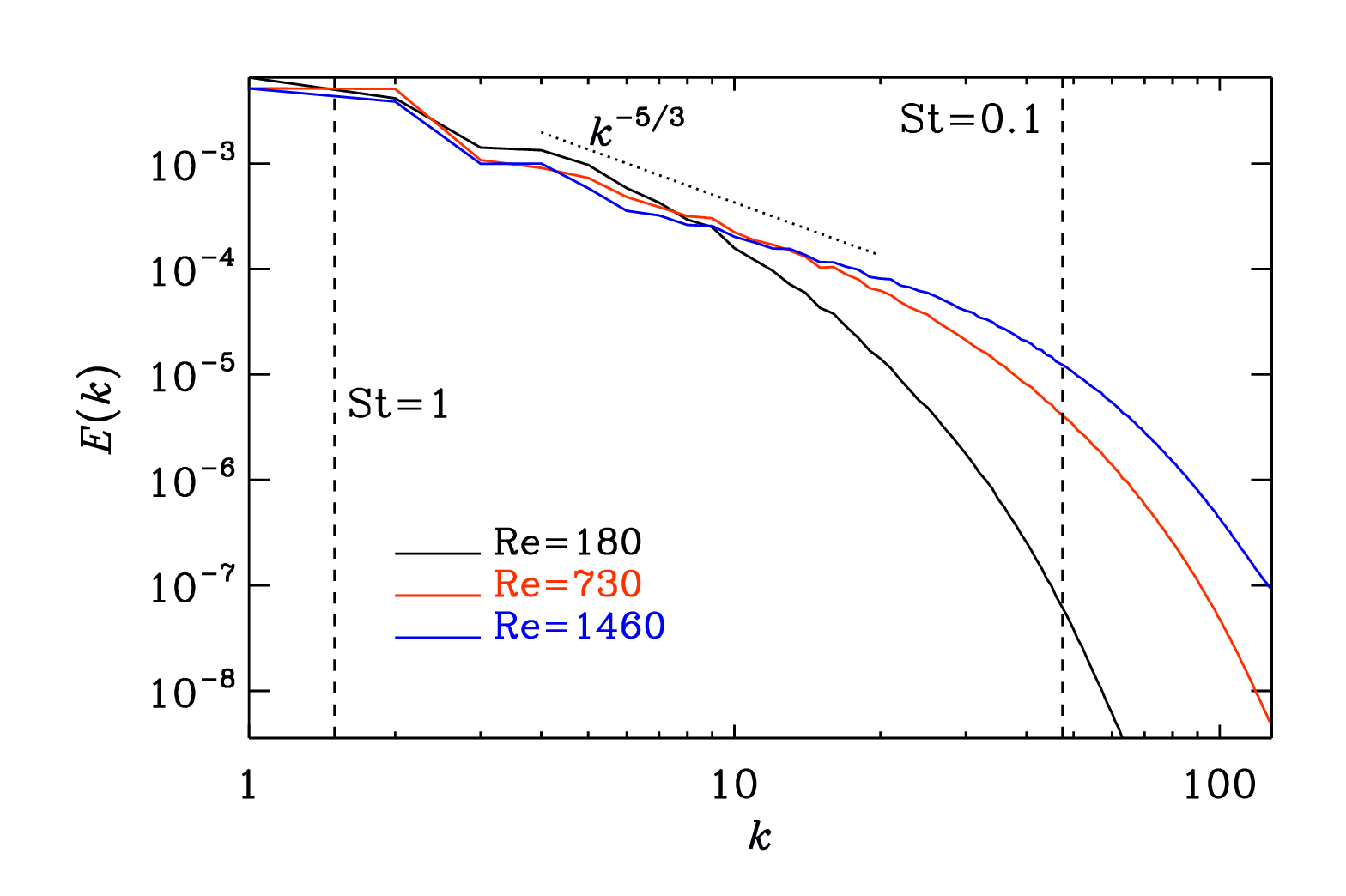}
\centering\includegraphics[width=.45\textwidth]{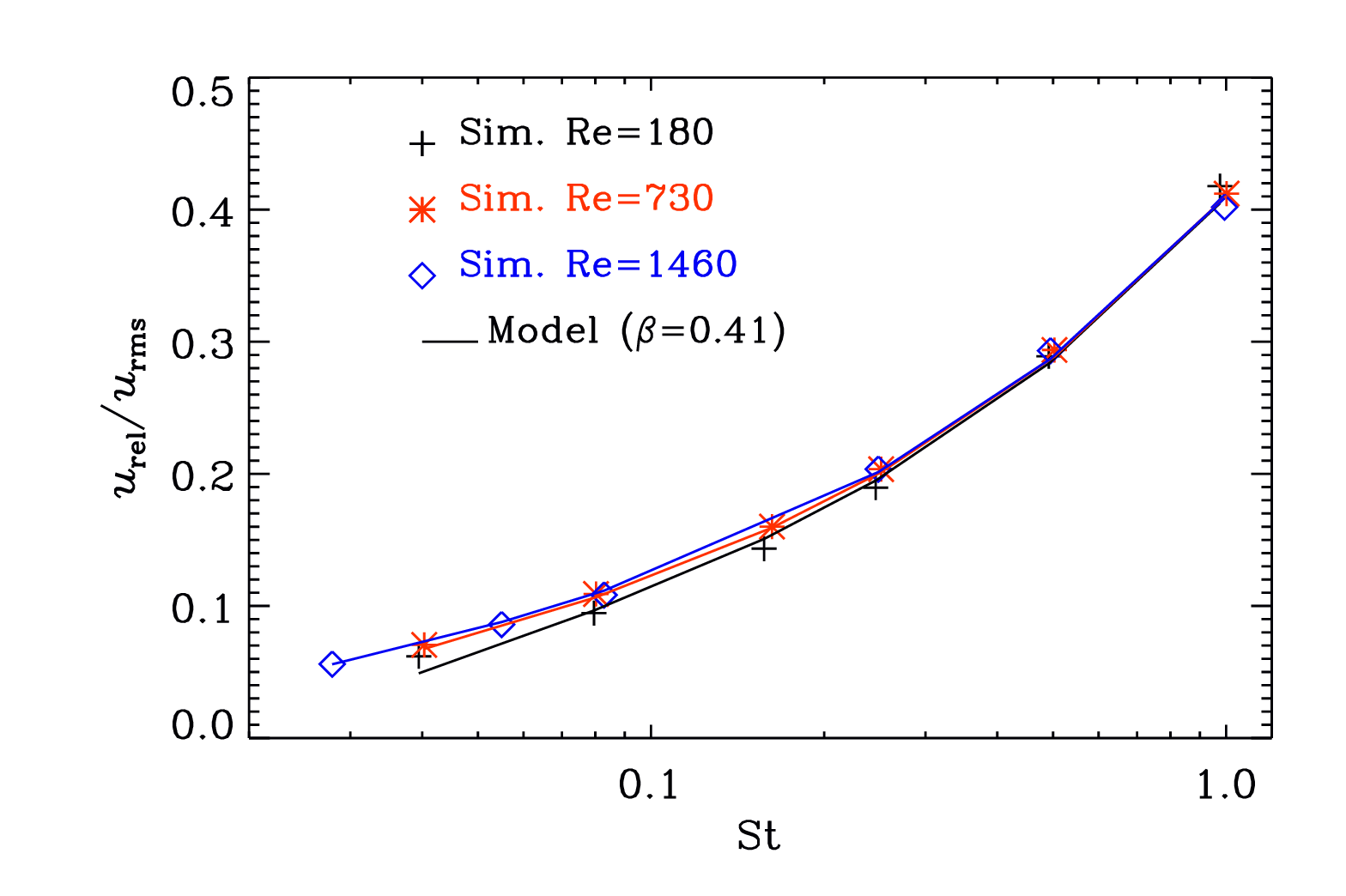}
\caption{Left panel: kinetic energy spectrum for different Reynolds
  numbers.  Right panel: relative particle velocity as a function of
  Stokes number.}
\label{fig:power_urel}
\end{figure}

Since all scales smaller than $\ell$ will induce a relative velocity between the
particles and the fluid, it is reasonable to assume that the 
relative velocity between
the fluid and the particles will be a certain fraction $\beta$ of the 
integrated turbulent velocity $\tilde{u}_\ell$
of all scales smaller than $\ell$, such that
\EQ
\label{urel0}
u_{\rm rel}=\beta \tilde{u}_\ell
\EN
when $\tilde{u}_\ell$ is defined as
\EQ
\label{utilde}
\frac{1}{2}\tilde{u}_\ell^2=\int_{k_\ell}^{k_\eta} E(k)dk
\EN
and $k_\eta=2\pi/\eta$ is the wave-number of the Kolmogorov scale
($\eta=(\nu^3/\epsilon)^{1/4}$), where $\epsilon$ is the dissipation 
rate of turbulent kinetic energy. Integration of \Eq{utilde} yields
\EQ
\label{utilde2}
\tilde{u}_\ell=u_{\rm rms} \sqrt{\frac{k_\ell^{-2/3}-k_\eta^{-2/3}}{k_L^{-2/3}-k_\eta^{-2/3}}}
\EN
for $E(k)=c\epsilon^{2/3}k^{-5/3}$ when it has been used that the 
total turbulent kinetic energy is given by
\EQ
\frac{1}{2}u_{\rm rms}^2=\int_{k_1}^{k_\eta} E(k)dk,
\EN
where $k_1$ is the wavenumber of the largest scale in the simulation.
Combining \Eqs{kl}{utilde2} with \Eq{urel0} finally yields
\EQ
\label{urel}
u_{\rm rel}=\beta u_{\rm rms}\sqrt{\frac{\St k_L^{-2/3}-k_\eta^{-2/3}}{k_L^{-2/3}-k_\eta^{-2/3}}}.
\EN
The unknown constant in this equation, $\beta$, 
can be determined numerically from 
\Eq{urel0}, i.e. $\beta=u_{\rm rel}/\tilde{u}_\ell$. Here, $u_{\rm rel}$ 
is found directly
from DNS simulations, while $\tilde{u}_\ell$ 
is calculated from \Eq{utilde2}.
It is seen from \Fig{fig:beta} that $\beta$ is close to 0.41 for most
Stokes and Reynolds numbers. The main exception is for low Reynolds
and Stokes numbers, where $\beta$ is significantly larger. This can 
be understood by inspecting the left panel of \Fig{fig:power_urel}, 
where it is seen that
for $\Rey=180$ and $\St<0.1$, we are already far into the dissipative 
subrange, where our model is not expected to be correct since it 
relies on a Kolmogorov scaling. 

It is surprising to see
that \Eq{urel} reproduces the relative particle velocity 
for such low Stokes numbers, even for the smaller
Reynolds numbers.
This may be explained by reconsidering \Eq{res_time}, where we 
assumed that the resonant eddies correspond to the eddies that
have {\it exactly} the same turnover time as the response time of the
particles. This is just an order of magnitude estimate, and a more
correct expression would probably be
\EQ
\label{res_time2}
\tau_\ell=\gamma\tau_p,
\EN
where $\gamma$ is of the order of unity.
More work should, however, be devoted to understanding the coupling between
the particles and the turbulent eddies. In particular, a more exact definition
of the resonant eddies is needed.
We nevertheless believe that
$\beta$ is a universal property of the HIP approximation and the Navier-Stokes
equations that
will have a constant value for all $\Rey$ and $\St$
as long as the resonant eddies are within the inertial range.

In the right panel of \Fig{fig:power_urel}, 
the average relative particle velocity, as found
from the DNS simulations (symbols), is compared with the predicted 
values from \Eq{urel} (solid lines). It is seen that the 
fit is rather good for most Reynolds and Stokes numbers. This
supports the use of \Eq{urel} for predicting the relative particle
velocity.

\subsection{The cluster size}
\begin{figure}
\centering\includegraphics[width=.8\textwidth]{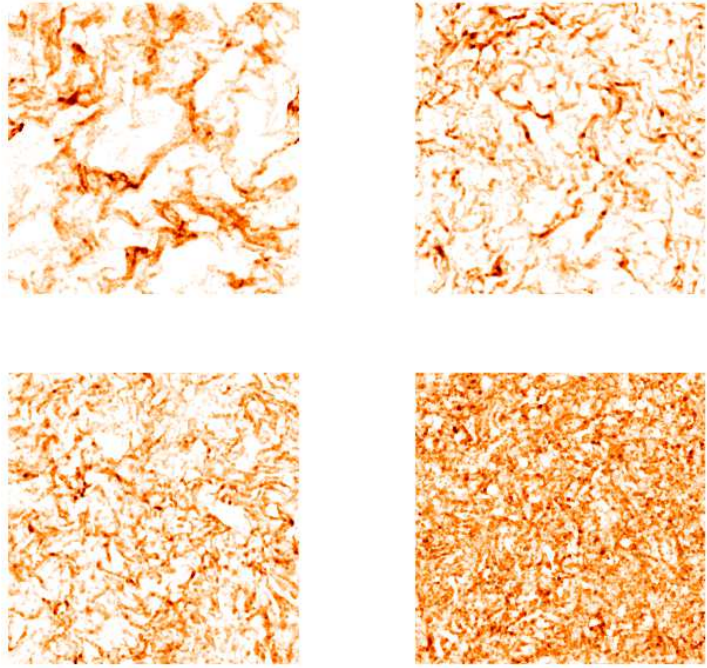}
\caption{Particle number density for $\St=1$ (upper left), 
$\St=0.3$ (upper right), $\St=0.1$ (lower left) and $\St=0.03$ 
(lower right) (runs 3A, 3B, 3C and 3D in \Tab{runs}).}\label{fig:box_snap}
\end{figure}

\begin{figure}
\centering\includegraphics[width=.5\textwidth]{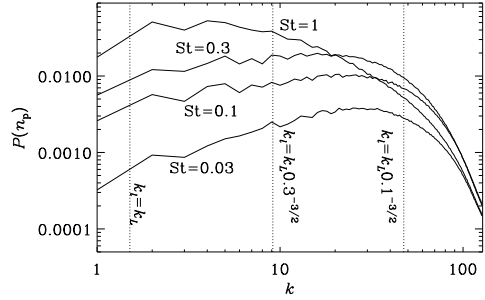}
\caption{Power spectrum of particle number density for
runs 3A, 3B, 3C and 3D in \Tab{runs}.}
\label{fig:box_power}
\end{figure}

The typical size of the clusters $\ell$ is assumed to be 
the size of the resonant eddies. From \Eq{kl} this
yields a cluster size of 
\EQ
\label{lL}
l=L\St^{3/2}.
\EN
It can be seen from \Fig{fig:box_snap} that the particle number density
distribution does indeed show more small scale variation for 
the smaller Stokes numbers. 
This has been quantified in \Fig{fig:box_power}
where the power spectrum of the particle number density is shown. Here
we see that the spectrum peaks at large scales for $\St=1$ while
the peak is located at much smaller scales for smaller Stokes numbers. 
The peak in the spectrum does not, however, follow \Eq{lL} as 
accurately as expected. The reason
for this is most likely that power spectra are not the right diagnostics
to study the size of particle clusters, but it may also be partly 
because of: 
1) poor statistics due to too few particles (the smaller
clusters are not filled with particles), 
2) the constant in the 
definition of the resonant eddies not being unity (see e.g. \Eq{res_time2}), or 
3) finite Reynolds number effects. 

The power spectrum $P$ can be integrated to yield 
a measure of the strength in the
particle number density fluctuations, given by the root-mean-square (rms) 
particle number density;
\EQ
n_{\rm rms}=\int P dk.
\EN  
It is found that the rms particle number density is decreasing with 
Stokes number. More specifically, $n_{\rm rms}$ is 1.6, 1.5, 1.2 and 0.8
for Stokes numbers of 1, 0.3, 0.1 and 0.03, respectively. This means that
the high density regimes have higher particle number densities for
larger Stokes numbers.

\subsection{Reactant consumption rate}
\begin{figure}
\centering\includegraphics[width=.48\textwidth]{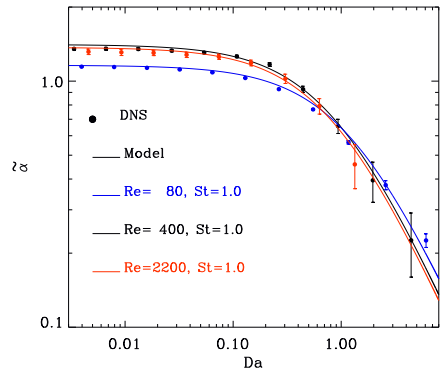}
\caption{Normalized decay rate as a function of \dam number 
for Stokes number of unity (runs 1A-3A).}
\label{fig:alpha}
\end{figure}
\begin{figure}
\centering\includegraphics[width=.48\textwidth]{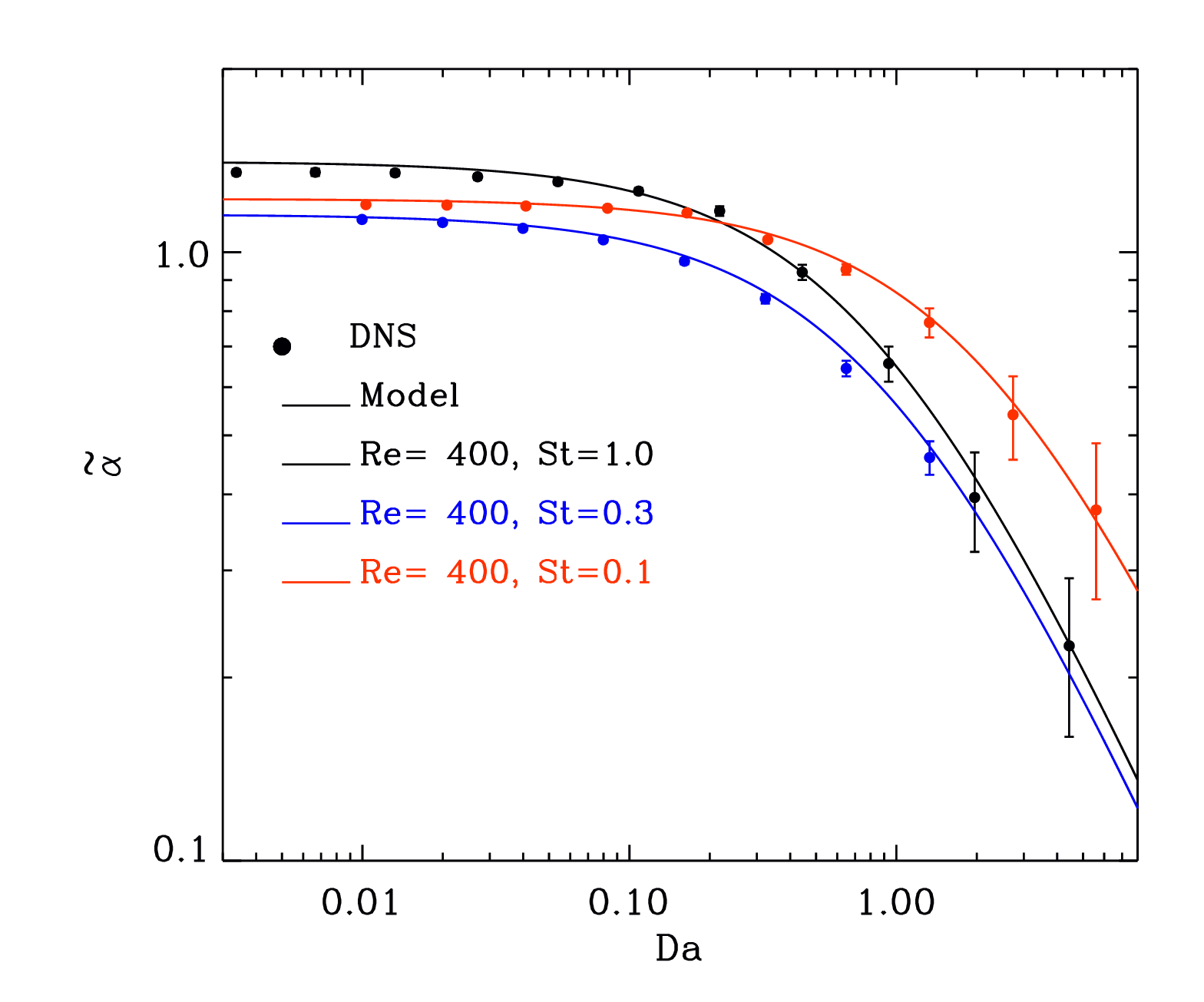}
\centering\includegraphics[width=.48\textwidth]{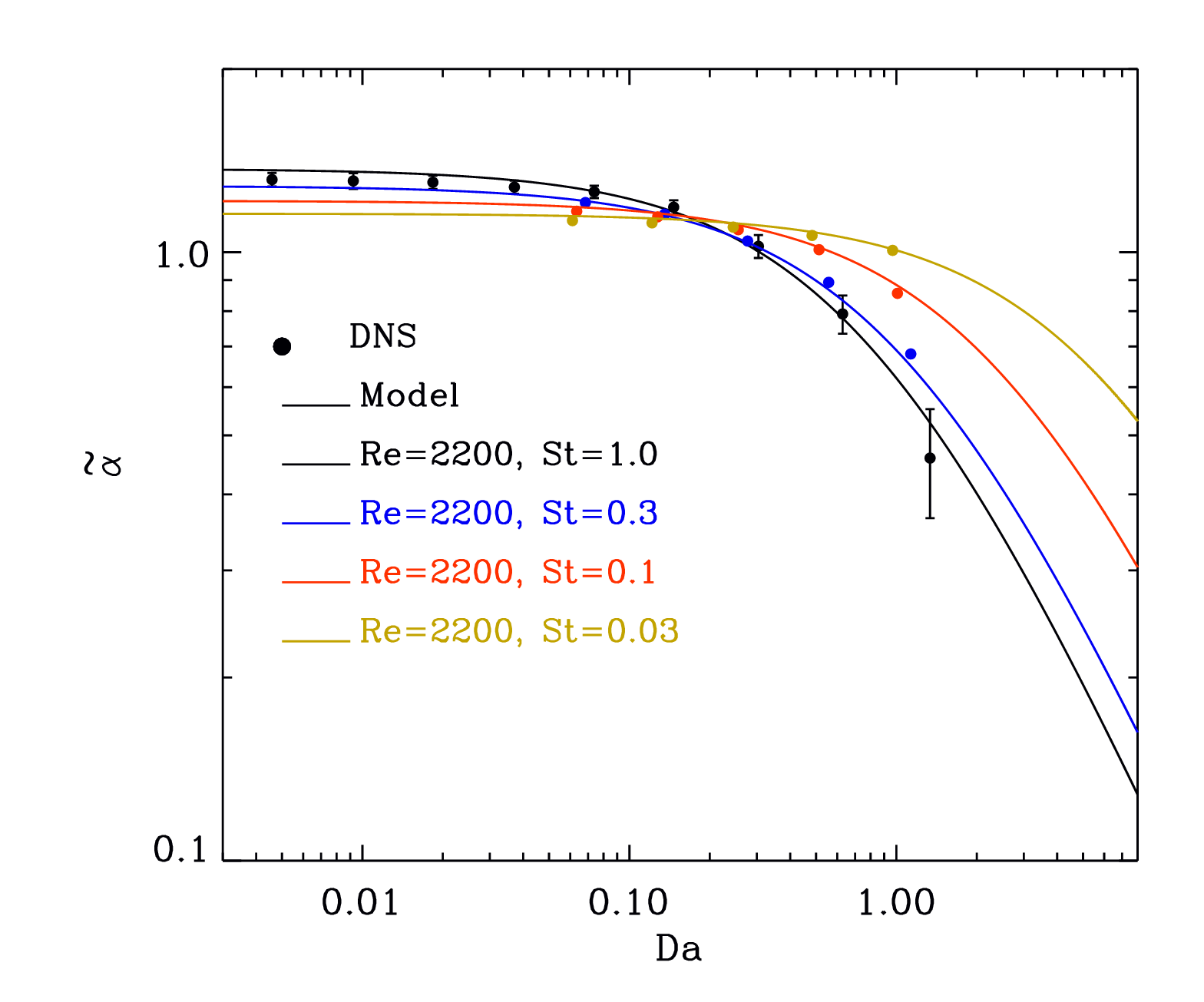}
\caption{Normalized decay rate as a function of \dam number 
for different Stokes numbers. The left panel show the results
for $\Rey=400$ (runs 2A-C) while the right panel is 
for $\Rey=2200$ (runs 3A-3D).}
\label{fig:alpha_St}
\end{figure}
The normalized reactant consumption rate is shown in \Fig{fig:alpha}. 
The symbols correspond to the results from
the DNS simulations, while the solid lines are given by \Eq{alpha_tilde}.
Here, the Stokes number is found by using the model for the relative 
velocity, as given by \Eq{urel}, in the expression for the
Sherwood number (\Eq{sherwood}). The value of the cluster decay rate, 
$\alpha_c$, is the only free parameter and it is chosen by a best fit approach.
The values of $\alpha_c$ are found in \Tab{runs}. 

The value of $\tilde{\alpha}$ for small \dam numbers equals the 
Sherwood number divided by two, while the \dam number for which $\tilde{\alpha}$
starts to decrease is determined by the cluster decay rate $\alpha_c$.
Overall, the model seems to follow the results from the DNS simulations 
rather well. 

From \Fig{fig:alpha} it can be seen  
that for large Stokes numbers,
the curves for the normalized decay rates of a given Stokes number
overlap for different Reynolds numbers if the Reynolds number is high 
enough. 
This is because the resonant eddies are at scales larger than
the dissipative subrange. So
increasing the Reynolds number, which may be considered a shift
of the dissipative subrange to smaller scales, is not affecting 
the resonant eddies, and hence also the clustering is unaffected.
If, however, the Reynolds or the Stokes number is small, such 
that the resonant eddies are in the
dissipative subrange, a change in Reynolds number will have an effect
on the normalized decay rate ($\tilde{\alpha}$).

Figure \ref{fig:alpha_St} shows that by decreasing
the Stokes number, the normalized reactant decay rate
stays unchanged up to larger \dam numbers. 
This means that the effect of particle clustering is weaker for
smaller Stokes numbers.
This is expected
since the limit of very small clusters corresponds to individual
particles, where $\tilde{\alpha}$ is independent of $\Da$.
From the simulations with small \dam numbers and $\Rey=2200$, 
which are shown in the right panel of \Fig{fig:alpha_St}, 
it can be observed that the
normalized decay rate is monotonically decreasing with Stokes number.
The reason for this is that for these simulations 
the particle size is kept constant as the Stokes
number is changed, such that 
the Sherwood number is decreased with decreasing Stokes number. 
This is, however, not the case for the simulations with
$\Rey=400$, where it is found that  
the normalized decay rate for small \dam numbers
is lower for $\St=0.3$ than for $\St=0.1$. The reason for this is 
that a smaller particle radius was used for the simulations with 
$\St=0.3$. The effect of reducing the particle radius is 
that the particle Reynolds number,
and hence also the Sherwood number, is decreased.

\subsection{The cluster decay rate}
If the chemical time scale is much shorter than the lifetime of 
the particle clusters, the interior of the clusters will quickly be void
of reactants. This means that the reactant consumption rate is
controlled by the flux of reactant to the surface of the clusters,
where the reactant will be consumed at the exterior
of the cluster. Based on this, it is clear  
that for large $\Da$ (small $\tau_c$), 
the clusters behave as large solid particles, or super-particles.
Following \cite{kruger}, 
the reactant decay rate is then given by the so called 
cluster decay rate;
\EQ
\alpha_c=n_c\kappa_cA_c
\EN
when $n_c=A_1l^{-3}$ is the number density of clusters (or super-particles),
$\kappa_c=D_t\Sh/l$ is the reactant diffusion rate to the 
super-particles, $A_c=A_2l^2$ is the surface area of the clusters,
$D_t$ is the turbulent diffusivity that carries the reactant from
the surrounding 
fluid to the surface of the clusters and $A_1$ and $A_2$ are
constants that depend on the dimensionality of the clusters.
It is clear that turbulent eddies larger than $\ell$, as given by 
\Eq{lL},
can not participate in the turbulent transport of reactants to the
clusters, while eddies slightly smaller than $\ell$ will participate.
A first approximation of the turbulent diffusivity to the surface 
of the clusters is therefore given by
\EQ
D_t=u_\ell l=u_LL\St^{2}.
\EN
By combining the above, taking into account \Eq{lL}, it can be found that
\EQ
\frac{\alpha_c\tau_L\St}{\Sh}=A_1A_2,
\EN
where the right hand side should be constant for resonant 
eddies well inside the inertial range.
\begin{figure}
\centering\includegraphics[width=.5\textwidth]{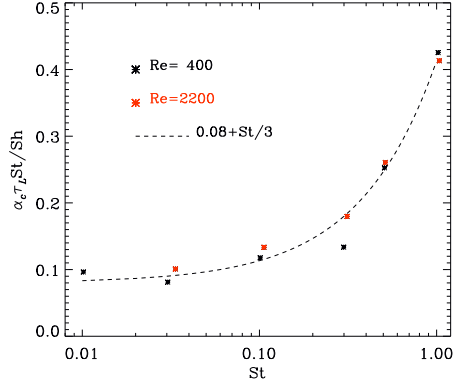}
\caption{The product $\alpha_c\tau_L\St$ as a function of $\St$ for
runs with resonant eddies in the inertial range.}
\label{fig:alphac}
\end{figure}
From \Fig{fig:alphac}, it can be seen that the right hand side
of the above equation is constant only
for Stokes numbers smaller than $\sim 0.3$. 
Since the value of the right hand side starts to increase
already for $\St=0.3$, this may
once again indicate that $\gamma$ from \Eq{res_time2}
is different from unity. 
The discrepancy may also be due to the fact that 
when it comes to the shape of the particle clusters,  
a large scale strain may stretch the particle clusters.
For $\St\sim 1$, there are no vortices that are larger than 
the clusters, and hence the dimensionality of the clusters
becomes different. This will inevitably yield different values 
of $A_1A_2$.
The value of the geometric coefficients can be fitted by
\EQ
A_1A_2=0.08 + \St/3,
\EN 
but this is just an empirical fit and more work is 
required in order to understand the fundamentals behind 
the shape and size of the particle clusters.

\section{Conclusion}
In this work, the effect of turbulence on the mass (and heat) transfer
between inertial particles and the embedding fluid is studied. The
turbulence is shown to have two effects on the mass transfer. The
first effect is active for all \dam numbers, and here the turbulence
increases the mass transfer rate due to the relative velocity between
the particles in the fluid. A corresponding model for the relative
velocity between the fluid and the particles is given by
\Eq{urel} which uses basic variables of the flow. With this, adding 
effects of relative velocity into RANS based simulations is possible. 

The second effect with which turbulence influences the mass
transfer rate is through the clustering of particles. It is shown
that the size of the particle clusters increases with the particle Stokes
number, and that the clustering decreases the overall mass transfer
rate between the particles and the fluid. 
This is a confirmation of the findings of \cite{kruger}.
In addition, a model is developed that
takes this effect into account and incorporates it into a modified
Sherwood number.  This model is shown to give reasonable results for
Stokes numbers (based on the turbulent integral scale) less than $\sim
0.3$, while an empirical fit is employed to account for Stokes
numbers up to unity. 
More work is still required in order to fully understand the
size and dimensionality of the the particle clusters. As of now, a
unique way of characterizing particle clusters does not exist, and
very little work has actually been put into the study of large-scale
clustering of particles.

\section*{Acknowledgements}
The research leading to these results has received funding from the
Polish-Norwegian Research Program operated by the National Centre
for Research and Development under the Norwegian Financial Mechanism
2009-2014 in the frame of Project Contract No Pol-Nor/232738/101/2014.
This work was supported by the grant "Bottlenecks for particle growth 
in turbulent 
aerosols” from the Knut and Alice Wallenberg Foundation, Dnr. KAW 2014.0048 and
by grant from Swedish Research Council (Dnr. 638-2013-9243).
NELH and DM also acknowledge the Research Council of
Norway under the FRINATEK grant 231444.

\bibliographystyle{jfm}

\begin{thebibliography}{32}

\bibitem[Jones \& Launder~(1972)]{jones_launder}
W. P. Jones and B. E. Launder\yjour{1972}{Int. J. Heat Mass Transfer}{15}{301}

\bibitem[Pope~(2003)]{pope}
S. B. Pope\ybook{2003}{Turbulent flows}{Cambridge University Press}

\bibitem[Schiller \& Naumann~(1933)]{schiller1933}
L. Schiller and A. Naumann\yjour{1933}{Ver. Deut. Ing}{77}{318}

\bibitem[Crowe~(2012)]{crowe2012}
C. T. Crowe\ybook{2012}{Multiphase Flows Droplets}{CRC Press}

\bibitem[Magnussen \& Hjertager(1976)]{EDM}
B. F. Magnussen and B. H. Hjertager\yypr{1976}{719-729}{16'th Symposium
(International) on Combustion}{Pittsburgh: The Combustion Institute}

\bibitem[Ertesv{\aa}g \& Magnussen~(2000)]{EDC}
I. S. Ertesv{\aa}g and B. F. Magnussen \yjour{2000}{Combust. Sci.
Technol.}{159}{213}

\bibitem[Dopazo~(1994)]{PDF}
C. Dopazo\yproc{1994}{375-474}{P. A. Libby and F. A. Williams}{London: Academic
Press.}

\bibitem[Ranz \& Marshall~(1952)]{ranz_marshall}
W. E. Ranz and W. R. Marshall\yjour{1952}{Chem Engr. Prog.}{48}{pps. 141 and
173}

\bibitem[Kruger et al.~(2016)]{kruger}
J. Kruger, N. E. L. Haugen, T. Lovas and D. Mitra {\it Proc. Comb. Symp.}, dx.doi.org/10.1016/j.proci.2016.06.187


\bibitem[Silaen \& Wang~(2010)]{silaen_wang_10}
A. Silaen and T. Wang\yjour{2010}{Int. J. Heat Mass Transfer}{53}{2074-2091}

\bibitem[Vascellari et al.~(2014)]{vascellari_etal_14}
M. Vascellari, S. Schulze, P. Nikrityuk, D. Safronov and C.
Hasse\yjour{2014}{Flow turbulence and combustion}{92}{319-345}

\bibitem[Vascellari et al.~(2015)]{vascellari_etal_15}
M. Vascellari, D. G. Roberts, S. S. Hla, D. J. Harris and C.
Hasse\yjour{2015}{Fuel}{152}{58-73}

\bibitem[Klimanek et al.~(2015)]{klimanek_etal_15}
A. Klimanek, W. Adamczyk, A. Katelbach-Wozniak, G. Wecel and A.
Szlek\yjour{2015}{Fuel}{152}{131-137}

\bibitem[Chen et al.~(2012)]{chen_etal_12}
L. Chen, S. Z. Yong and A. F. Ghoniem\yjour{2012}{Progress in energy and
combustion science}{38}{156-214}

\bibitem[Chen et al.~(2000)]{chen_etal_00}
C. Chen, M. Horio and T. Kojima\yjour{2000}{Chemical Engineering
Science}{55}{3875-3883}

\bibitem[Gao et al.~(2004)]{gao_2004}
J. Gao, C. Xu. S. Lin and G. Yang\yjour{2004}{AIChE Journal}{45}{1095-1113}

\bibitem[Zhang et al.~(2005)]{zhang_2005}
Y. Zhang, X.-L. Wei, L.-X. Zhou and H.-Z. Sheng\yjour{2005}{Fuel}{84}{1798-1804}

\bibitem[Luo et al.~(2012)]{luo_2012}
K. Luo, H. Wang, J. Fan and F. Yi\yjour{2012}{Energy and Fuels}{26}{6128-6136}

\bibitem[Brosh \& Chakraborty~(2014)]{brosh_2014}
T. Brosh and N. Chakraborty\yjour{2014}{Energy and Fuels}{28}{6077-0688}

\bibitem[Brosh et al.~(2015)]{brosh_2015}
T. Brosh, D. Patel, D. Wacks and N. Chakraborty\yjour{2015}{Fuel}{145}{50-62}

\bibitem[Hara et al.~(2015)]{hara_2015}
T. Hara, M. Muto, T. Kitano, R. Kurose and S. Komori\yjour{2015}{Combustion and Flame}{162}{4391-4407}

\bibitem[Deen \& Kuipers~(2014)]{deen_kuipers_2014}
N. G. Deen and J. A. M. Kuipers\yjour{2014}{Chemical Engineering Science}{116}{645-656}

\bibitem[Johansen et al.~(2007)]{johansen07}
A. Johansen, J. S. Oishi, M.M.Mac Low, H. Klahr, T. Henning, and A.
Youdin\yjour{2007}{Nature}{448}{1022}

\bibitem[Haugen \& Kragset~(2010)]{haugen_kragset10}
N. E. L. Haugen and S. Kragset\yjour{2010}{J. Fluid Mech.}{661}{239}

\bibitem[Haugen et al.~(2012)]{haugen_etal12}
N. E. L. Haugen, N. Kleeoring, I. Rogachevskii and A.
Brandenburg\yjour{2012}{Phys. Fluids}{24}{075106}

\bibitem[The Pencil-Code]{PC}
The Pencil-Code, http://pencil-code.nordita.org/

\bibitem[Haugen \& Brandenburg~(2006)]{haugen_brandenburg06}
N. E. L. Haugen and A. Brandenburg\yjour{2006}{Phys. Fluids}{18}{075106}

\bibitem[Calzavarini et al.~(2008)]{toschi}
E. Calzavarini, M. Kerscher, D. Lohse and F. Toschi\yjour{2008}{J. Fluid Mech.}{607}{13-24}

\bibitem[Baum \& Street~(1971)]{baum}
M. M. Baum and P. J. Street\yjour{1971}{Combustion Science and Technology}{3}{231-243}

\bibitem[Bec et al.~(2007)]{bec07}
J. Bec, L. Biferale, M. Cencini, A. Lanotte, S. Musacchio and F. Toschi\yjour{2007}{Physical Review Letter}{98}{084502}

\bibitem[Squires \& Eaton~(1991)]{Squires1991}
K. D. Squires and J. K. Eaton\yjour{1991}{Journal of Fluid Mechanics}{226}{1-35}

\bibitem[Wood et al.~(2005)]{Wood2005}
A. M. Wood, W. Hwang and J. K. Eaton\yjour{2005}{International Journal of Multiphase Flow}{31}{1220-1230}

\bibitem[Eaton \& Fessler~(1994)]{Eaton1994}
J. K. Eaton and J. R. Fessler\yjour{1994}{International Journal of Multiphase Flow}{20}{169-209}

\bibitem[Toschi \& Bodenschatz~(2009)]{Toschi2009}
F. Toschi and E. Bodenschatz\yjour{2009}{Annual Review of Fluid Mechanics}{41}{375-404}

\end{thebibliography}
{}

\end{document}